\documentclass[%
 reprint,
 superscriptaddress,
 amsmath,amssymb,
 aps,
 pra,
floatfix,
showkeys
]{revtex4-1}

\usepackage{graphicx}
\usepackage{dcolumn}
\usepackage{bm}
\usepackage{hyperref}
\hypersetup{colorlinks=true, citecolor=blue, urlcolor=black, linkcolor=blue}

\begin{document}


\title{\texorpdfstring{Bright Optical Eigenmode with about 1 nm$^3$ Volume}{Bright Optical Eigenmode with about 1 nm3 Volume}}

\author{Wancong Li}
\affiliation{School of Physics, Wuhan National Laboratory for Optoelectronics, Huazhong University of Science and Technology, Luoyu Road 1037, Wuhan, 430074, People's Republic of China}

\author{Qiang Zhou}
\affiliation{School of Physics, Wuhan National Laboratory for Optoelectronics, Huazhong University of Science and Technology, Luoyu Road 1037, Wuhan, 430074, People's Republic of China}

\author{Pu Zhang}
\email[Corresponding author, Email: ]{puzhang0702@hust.edu.cn}
\affiliation{School of Physics, Wuhan National Laboratory for Optoelectronics, Huazhong University of Science and Technology, Luoyu Road 1037, Wuhan, 430074, People's Republic of China}

\author{Xue-Wen Chen}
\email[Corresponding author, Email: ]{xuewen\_chen@hust.edu.cn}
\affiliation{School of Physics, Wuhan National Laboratory for Optoelectronics, Huazhong University of Science and Technology, Luoyu Road 1037, Wuhan, 430074, People's Republic of China}



\begin{abstract}
    Concentrating optical field in an eigenmode with a tiny volume is vitally important for light-matter interactions at the fundamental level and underpins new technologies. In the past decades, researchers have investigated various approaches to shrink light and so far managed to reduce the volume of optical eigenmodes, under the quantum-optical definition, down to several ten nm$^3$. Here we report on the discovery, characterization and engineering of a class of extremely localized eigenmodes that are resonances of atomistic protrusions on a metallic host nanoparticle and feature quantum-optical mode volumes of below 1 nm$^3$. We theoretically demonstrate that these extremely localized modes can be made bright with radiation efficiencies reaching $30\%$ and provide up to $4{\times}10^7$ times intensity enhancement. The existence of bright eigenmodes with the volume comparable to a photon emitter foresees exciting new optical physics, such as ultrastrong coupling with single optical emitters, angstrom-resolution optical imaging, and atomic-scale single-molecule photochemistry. 
\end{abstract}

\maketitle


\section{Introduction}

Shrinking optical photons towards the scale of electron wave function in an atom or molecule is the holy grail of nanophotonics, with new fundamental science and technologies envisioned. 
Owing to the law of diffraction, the confinement of photons is limited to ${\sim}(\lambda/2)^3$ with $\lambda$ being the wavelength of light. 
Over the past decades, the field of nanophotonics has flourished along with the development of various new concepts and techniques to overcome the diffraction limit\cite{Ebbesen2003SurfacePlasmons,Brongersma2010Plasmonics,Polman2015Nanophotonics}. 
Metallic nanostructures have been widely pursued for concentrating light into nanoscale volumes by mixing fields with conduction electrons, for instance, resonantly, via optical antennas\cite{Hulst2011Antennas} and metal-insulator-metal nanocavities\cite{Kurokawa2006Squeezing}, or non-resonantly, through the lightning rod effect at a sharp metal termination\cite{Nitzan1980ElectromagneticTheory}, nanofocusing with metal tapers\cite{Stockman2004Nanofocusing}. 
With resonant and non-resonant effects combined\cite{Brongersma2010Plasmonics,Ahmed2018ReachingLimits}, ultrathin metallic gaps have been actively studied for extreme light concentration\cite{Hecht2012AtomicScale,Cirac2012Probing,Baumberg2016NatureSingleMolecule,Baumberg2019ExtremeNanophotonics,Frank2020FarFieldExcitation}with mode volumes reduced down to several hundred nm$^3$, under the quantum-optical definition which characterizes light-matter interaction strength together with the mode quality factor\cite{ExploringQuantum,Koenderink2010PurcellFactor,Kristensen2014ModeVolume}. 
More recently, researchers have investigated experimentally\cite{Baumberg2016ScienceSingleMolecule,Dong2020photoluminescence} and theoretically\cite{Aizpurua2015Atomistic,Aizpurua2015classical,Aizpurua2018AtomicScale} individual atomistic features inside an ultrathin gap for further field localization.
The atomistic protrusions could naturally occur on metallic host nanoparticles as evident from transmission electron microscopy measurements\cite{Aizpurua2015classical,Takayanagi1998Qauntized} or be controllably assembled to the host\cite{Dong2020photoluminescence}.
The contribution of the atomistic protrusion was recognized as the non-resonant lightning rod effect on top of the gap-plasmon mode. 
Theoretical calculations\cite{Baumberg2016ScienceSingleMolecule} show such modified gap-plasmon modes have the quantum-optical mode volumes of several ten nm$^3$, still orders of magnitude larger than the typical size of an optical emitter.

Recent tip-enhanced optical imaging experiments have demonstrated spectacular spatial resolution, including sub-nanometer-resolution Raman mapping\cite{Zhang2013ChemicalMapping} and single-molecule photoluminescence\cite{Dong2020photoluminescence}, and angstrom-scale-resolution visualization of vibrational normal modes\cite{Apkarian2019vibrationalmodes}. 
These demonstrations unambiguously suggest that the optical field can be confined within a few angstroms in diameter and simultaneously possess huge local enhancement\cite{Dong2020photoluminescence}.
The physics behind these observations is, however, beyond the current understanding.
Especially, Landau damping (LD) of plasmons in metal, which accounts for single-electron excitation by large-wavevector field components\cite{Khurgin2017LandauDamping,Mortensen2015Nonlocal}, imposes a formidable challenge in achieving extreme localization and ultrastrong enhancement.
Here we report theoretically on the discovery, characterization and engineering of optical eigenmodes with the quantum-optical mode volumes down to $0.5$ nm$^3$. 
Such extremely localized modes (ELMs) originate from resonant accumulation of induced conduction electrons and currents around the atomic protrusions on a host metallic nanoparticle even though there is no inner physical boundary between the protrusion and host. 
Moreover, we present a strategy to overcome LD and reconcile the longtime dilemma of extreme localization and efficient radiation. Thus the ELMs can be made bright and readily accessible from the far field with intensity enhancement up to $4{\times}10^7$ folds.
The scenario of having an optical mode with volume about 1 nm$^3$ puts photons on the equal footing with an optical emitter, venturing into new regimes of optical physics.

\section{Theoretical foundation}
A rigorous modal analysis of the extremely localized field is indispensable to understand its mode structure and enables quantitative evaluation of the localization.
The inherent openness of the modes and absorption of metal require a modal analysis in the framework of quasi-normal modes (QNMs)\cite{Kristensen2014ModeVolume,Lalanne2018LigntInteraction,Philip2020biorthogonalQNM}.
However, modal analysis for a host nanoparticle with an atomistic protrusion is challenging because it possesses a relatively large host ($\sim$ $100$ nm) and an atomistic feature that is expected to exhibit quantum and many-body effects of conduction electrons, such as nonlocal response\cite{Cirac2012Probing,Dionne2012QuantumPlasmon}, electron spillover\cite{Hongxing2015ResonanceShifts} and LD\cite{Khurgin2017LandauDamping}.
Although \textit{ab initio} methods or time-dependent density functional theory (TD-DFT) can correctly describe these effects\cite{Aizpurua2015Atomistic,Zuloaga2009QuantumDescription}, they become computationally intractable for nanoparticles larger than a few nanometers.
Quantum hydrodynamic models (QHDMs) can mitigate the issue and describe microscopic details of macroscopic systems\cite{Hongxing2015ResonanceShifts,Mortensen2014nonlocal,Ding2017Plasmonic,cirac2017CurrentDependent}.
Notably, recent results of QHDM with appropriate electron kinetic energy approximations and a viscous stress term agree well with TD-DFT calculations and reproduce the size-dependent damping rate due to LD\cite{Khurgin2017LandauDamping} for nanospheres down to 1 nm in diameter\cite{cirac2017CurrentDependent}.
Here we develop a rigorous QNM method with the incorporation of conduction electron density and current in metal described by self-consistent QHDM. 

The conduction electrons are characterized by the number density $n(\mathbf{r},t)$ and velocity $\mathbf{v}(\mathbf{r},t)$. 
Assuming weak light excitation, the response of conduction electrons can be treated perturbatively, i.e., $n=n_0+n_1$, where $n_1$ is the light-induced dynamic electron density and $n_0$ the stationary density in absence of light excitation. 
The light-driven electrons lead to a current density $\mathbf{J}\approx n_0 q_\mathrm{e}\mathbf{v}$, with $q_\mathrm{e}$ being the electron charge. 
Under the perturbative treatment, a linearized hydrodynamic equation in frequency domain can be obtained as
\begin{equation}
\label{Eq:1}
(i\widetilde{\omega}+\gamma_\mathrm{D})\mathbf{J}
= \frac{q_\mathrm{e}}{m_\mathrm{e}}\left[
n_0q_\mathrm{e}\mathbf{E}
- n_0\nabla{\left(\frac{\delta{G}}{\delta{n}}\right)_1}
+ \nabla\cdot\overleftrightarrow{\sigma}
\right],
\end{equation}
where $m_\mathrm{e}$ and $\gamma_\mathrm{D}$ denote the electron mass and decay rate, respectively. 
The second term in the bracket includes the internal energy functional of the electron liquid $G=\int{d}\mathbf{r}\,g(n)$. The energy density $g(n)=T_\mathrm{TF}+T_\mathrm{w}+e_\mathrm{XC}$ consists of Thomas-Fermi kinetic energy $T_\mathrm{TF}$, von Weizs{\"a}cker kinetic energy $T_\mathrm{w}$ and exchange-correlation $e_\mathrm{XC}$. 
$(\cdots)_1$ means taking the terms linear to $n_1$. 
Here $\overleftrightarrow{\sigma}$ is the viscous stress tensor introduced to describe LD\cite{cirac2017CurrentDependent}. 
See Supplementary Note 1 for the explicit expressions of the above terms. 
The light-driven conduction current then serves as a secondary source for the optical field
\begin{equation}
\label{Eq:2}
i\widetilde{\omega}\,\varepsilon_\mathrm{b}(\widetilde{\omega})\,\mathbf{E}
= \nabla\times\mathbf{H}
- \mathbf{J} - \mathbf{J_s}.
\end{equation}
Here $\mathbf{J_s}$ is the external source current, and $\varepsilon_\mathrm{b}(\widetilde{\omega})$ is the background permittivity of the metal which may include the contribution from bound electrons for noble metals like Au (see Methods). 
We develop a QHDM-based QNM theory by incorporating equations (\ref{Eq:1}) and (\ref{Eq:2}) in the formulation\cite{Lalanne2018LigntInteraction} (Supplementary Note 1 presents a comprehensive introduction to the QHDM-based QNM method).
Since QNMs are self-sustained excitations of a linear open system with poles on the complex frequency plane as the eigen-frequencies, we employ an iterative procedure to locate the complex poles $\tilde{\omega}_q=\omega_q+i\gamma_q$ and obtain the corresponding mode profiles $\tilde{\mathbf{E}}_q$. 
To perform reliable QHDM and QNM calculations, we apply an efficient finite element method based implementation with multiscale meshing (see Methods). 
The quality factor and resonant wavelength of the QNM are given by $Q_q=\omega_q/(2\gamma_q)$ and $\lambda_q=2\pi{c}/\omega_q$ with $c$ being the light speed in vacuum, respectively. 
Under the quantum-optical definition\cite{ExploringQuantum,Marquier2017Revisiting}, the mode volume of a QNM reads
\begin{equation}
\label{Eq:3}
V_q(\mathbf{r})
= \frac{3Q_q\lambda_q^3}{4\pi^2n_\mathrm{d}^2(\mathbf{r})f_{\mathrm{P},q}(\mathbf{r})},
\end{equation}
where $f_{\mathrm{P},q}(\mathbf{r})$ is the Purcell factor contributed by this QNM and $n_\mathrm{d}(\mathbf{r})$ is the refractive index of dielectric medium. 
The mode volume in equation \eqref{Eq:3} is position dependent and we characterize the level of mode localization by its minimum 
$V_\mathrm{m}=|{\mathbf{\widetilde{E}}_q(\mathbf{r})}/{\mathbf{\widetilde{E}}_q(\mathbf{r}_\mathrm{M})}|^2\,V_q(\mathbf{r})$ at $\mathbf{r}_\mathrm{M}$, 
where $|\mathbf{\widetilde{E}}_q|$ reaches maximum\cite{Koenderink2010PurcellFactor,Kewes2018StrongCoupling}. 
To obtain $f_{\mathrm{P},q}(\mathbf{r})$, a dipole source $\mathbf{J_s}$ is applied to excite the system. 
The resulting electric field at the source position is expanded according to the Riesz projection procedure\cite{Binkowski2018RieszProjection} as $\mathbf{E}(\omega)=\sum_l\mathbf{E}_l(\omega)+\mathbf{E}_\mathrm{nr}$, where the sum accounts for the contributions of all QNMs involved and $\mathbf{E}_\mathrm{nr}$ is the remaining non-resonant contribution. 
Then one obtains
\begin{equation}
\label{Eq:4}
f_{\mathrm{P},q}
= -\frac{1}{2}\mathrm{Re}\left[\mathbf{E}_q(\omega_q)\cdot\mathbf{J_s^*}\right]/P_0,
\end{equation}
where $P_0$ is the radiation power of $\mathbf{J_s}$ in vacuum.

\section{Results}

\begin{figure*}[htbp]
\centering
\includegraphics[width=0.9\linewidth]{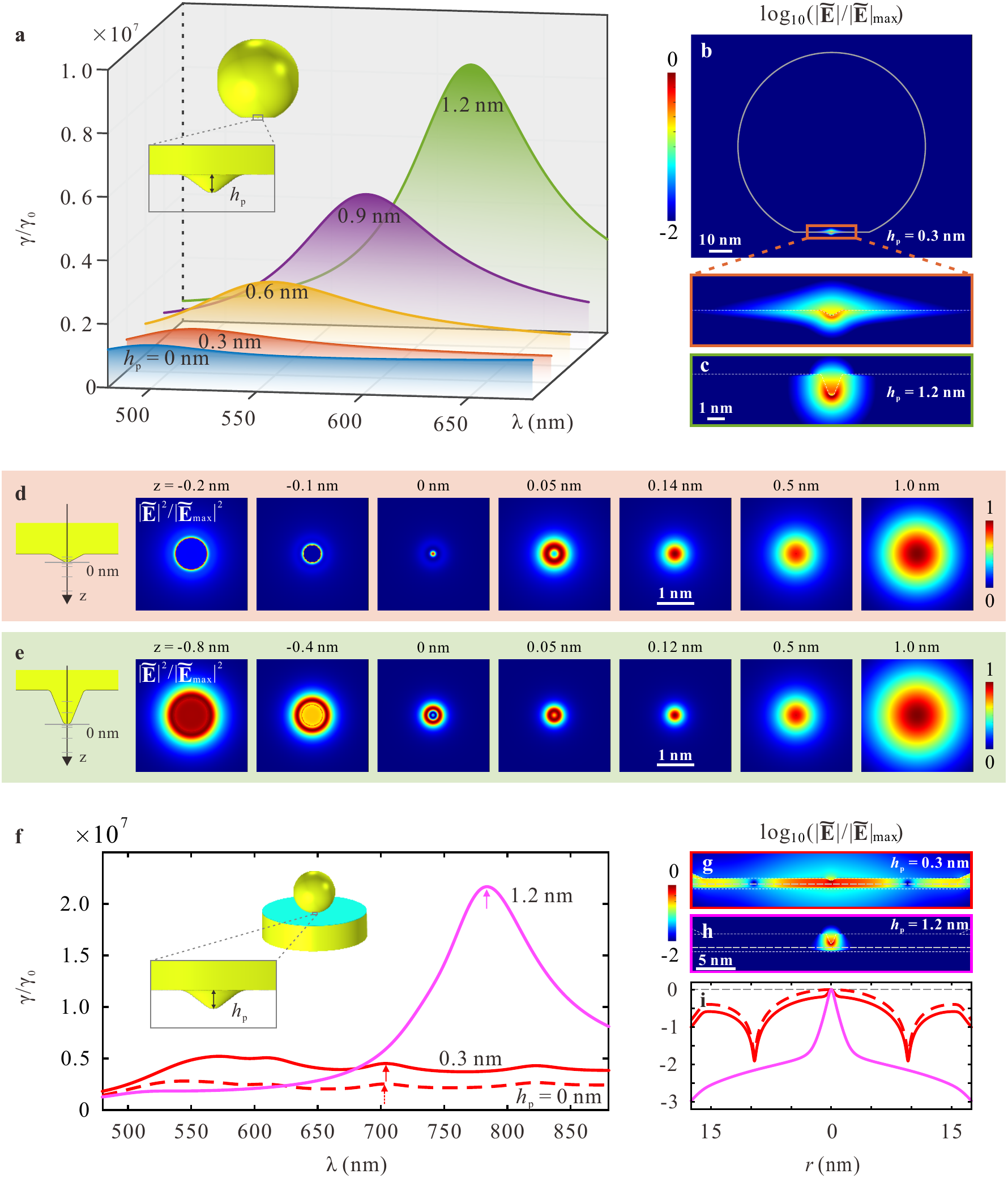}
\caption{\textbf{Extremely localized mode (ELM) near an atomistic protrusion of a host nanoparticle.} \textbf{a,} Spectra of emission enhancement for a dipole placed $0.5$ nm below a cone-shape protrusion tip shown in the inset and the zoom-in view. The protrusion has a fixed base diameter of $1.2$ nm and a varying height $h_\mathrm{p}$. \textbf{b,} ELM field profile in logarithmic scale for the $h_\mathrm{p} = 0.3$ nm case. \textbf{c,} ELM field profile near the protrusion for the $h_\mathrm{p} = 1.2$ nm case. \textbf{d} and \textbf{e,} Normalized intensity distributions at various distances from the tip with $h_\mathrm{p} = 0.3$ and $1.2$ nm, respectively. \textbf{f,} Spectra of emission enhancement for three NPoM structures shown in the inset. The dielectric spacer has a refractive index of $1.45$. \textbf{g} and \textbf{h} show the mode profiles in logarithmic scale for the resonances denoted by the arrows in \textbf{f} for $h_\mathrm{p} = 0.3$ and $1.2$ nm, respectively. \textbf{i,} Field profiles along the line $0.5$ nm above the substrate for all three denoted resonances..}
\label{Fig:1}
\end{figure*}

We begin quantitative discussion with a simple structure as shown in the inset of Fig. \ref{Fig:1}\textbf{a}, i.e., an $80$ nm diameter gold nanosphere with a $32$ nm diameter flat bottom, which hosts a cone-shape atomistic protrusion as depicted in the zoomed-in view. 
The protrusions under study have the same base diameter of $D_\mathrm{p}=1.2$ nm and a varying height of $h_\mathrm{p}=0.3$, $0.6$, $0.9$ and $1.2$ nm.
Complete geometry specifications are indicated in Supplementary Fig. 3.
To pinpoint the modal properties, we calculate the emission enhancement of a vertically oriented dipole source placed $0.5$ nm below the protrusion tip (such dipole position is used throughout the paper). 
The color-coded traces in Fig. \ref{Fig:1}\textbf{a} plot the spectra of emission enhancement for the above four structures and the case without a protrusion. 
With the introduction of the protrusion, a prominent resonance appears on top of the response of the host nanosphere and shows redshift as the protrusion height increases. 
The huge emission enhancement peaks (approaching $10^7$) imply the existence of a class of ELMs, whose resonant wavelength doesn’t depend on the host particle but instead on the protrusion shape (Supplementary Note 2). 
We perform QHDM-based QNM analysis to identify and characterize the ELMs. 
For the case with the bluntest protrusion ($h_\mathrm{p}=0.3$ nm), we find a complex-wavelength pole at $\widetilde{\lambda}=(489-43i)$ nm contributing the vast majority ($\sim80\%$) of the emission enhancement. 
Figure \ref{Fig:1}\textbf{b} displays the mode profile in logarithmic scale. 
The lower panel shows a zoomed-in view of the field near the protrusion. 
The other three protrusions all support such type of modes with the modal field extremely localized around the protrusion (Supplementary Note 2). 
Figure \ref{Fig:1}\textbf{c} depicts the ELM near the $h_\mathrm{p}=1.2$ nm protrusion. 
Fine features of the mode profiles are revealed by showing the normalized intensity distribution at various distances from the protrusion tip in Fig. \ref{Fig:1}\textbf{d} and  \ref{Fig:1}\textbf{e} for $h_\mathrm{p}=0.3$ and $1.2$ nm, respectively. 
The profiles for $h_\mathrm{p}=0.3$ nm are reminiscent of surface plasmon with the maximum appearing on the metal surface whereas for $h_\mathrm{p}=1.2$ nm the maximum lies about $0.12$ nm away from the metal. 
Outside the metal, the smallest spot sizes (full widths at half maximum) are $0.54$ nm at $z=0.14$ nm and $0.38$ nm at $z=0.12$ nm for $h_\mathrm{p}=0.3$ nm and $1.2$ nm, respectively. 
Our calculations show that the mode volumes for $h_\mathrm{p}=0.3$, $0.6$, $0.9$ and $1.2$ nm are $2.7$, $1.1$, $0.70$ and $0.59$ nm$^3$, respectively.

Having demonstrated the existence of ELMs on a standalone nanoparticle, we examine the nanoparticle-on-mirror (NPoM) structure that recently attracts considerable interest\cite{Cirac2012Probing,Baumberg2016NatureSingleMolecule,Baumberg2019ExtremeNanophotonics,Baumberg2016ScienceSingleMolecule}. 
As shown in the inset of Fig. \ref{Fig:1}\textbf{f}, we study the cases without a protrusion, with the $h_\mathrm{p}=0.3$ and $1.2$ nm protrusions in the dielectric nanogaps of sizes $1.3$, $1.3$ and $2.2$ nm, respectively. 
Figure \ref{Fig:1}\textbf{f} summarizes the spectra of emission enhancement for the three cases. 
The spectra of the NPoM without a protrusion and with the protrusion of $h_\mathrm{p}=0.3$ nm are quite flat due to the effect of LD\cite{Khurgin2017LandauDamping}. 
The recognizable resonance around $700$ nm for $h_\mathrm{p}=0.3$ nm was called the “picocavity” mode\cite{Baumberg2016ScienceSingleMolecule}, which is stronger than but similar in nature to the quadrupolar gap-plasmon resonance of the no-protrusion case (dashed line). 
For $h_\mathrm{p}=1.2$ nm, the situation is very different with a huge enhancement peak around $780$ nm. 
The resonance shift compared to the case in Fig. \ref{Fig:1}\textbf{a} is due to the dielectric spacer and the gap (Supplementary Note 3).
The corresponding mode profiles for $h_\mathrm{p}=0.3$, $1.2$ nm near the protrusion are displayed in logarithmic scale in Fig. \ref{Fig:1}\textbf{g} and \ref{Fig:1}\textbf{h}, respectively. 
Figure. \ref{Fig:1}\textbf{i} plots the field profiles along the line $0.5$ nm above the substrate for all three cases. 
While the field profile for $h_\mathrm{p}=0.3$ nm largely follows the profile of the no-protrusion case except a small bump due to the lightning rod effect from the protrusion, the ELM for $h_\mathrm{p}=1.2$ nm is clearly much more localized. 
The calculated mode volumes are $310$, $74$ and $0.47$ nm$^3$ for the no-protrusion, $h_\mathrm{p}=0.3$ and $1.2$ nm cases, respectively. 
From the examples shown in Fig. \ref{Fig:1}, one sees that the ELM exists without the need of an ultrathin gap and is completely different from the lightning-rod effect enhanced gap-plasmon mode.

\begin{figure}[htpb]
\centering
\includegraphics[width=\linewidth]{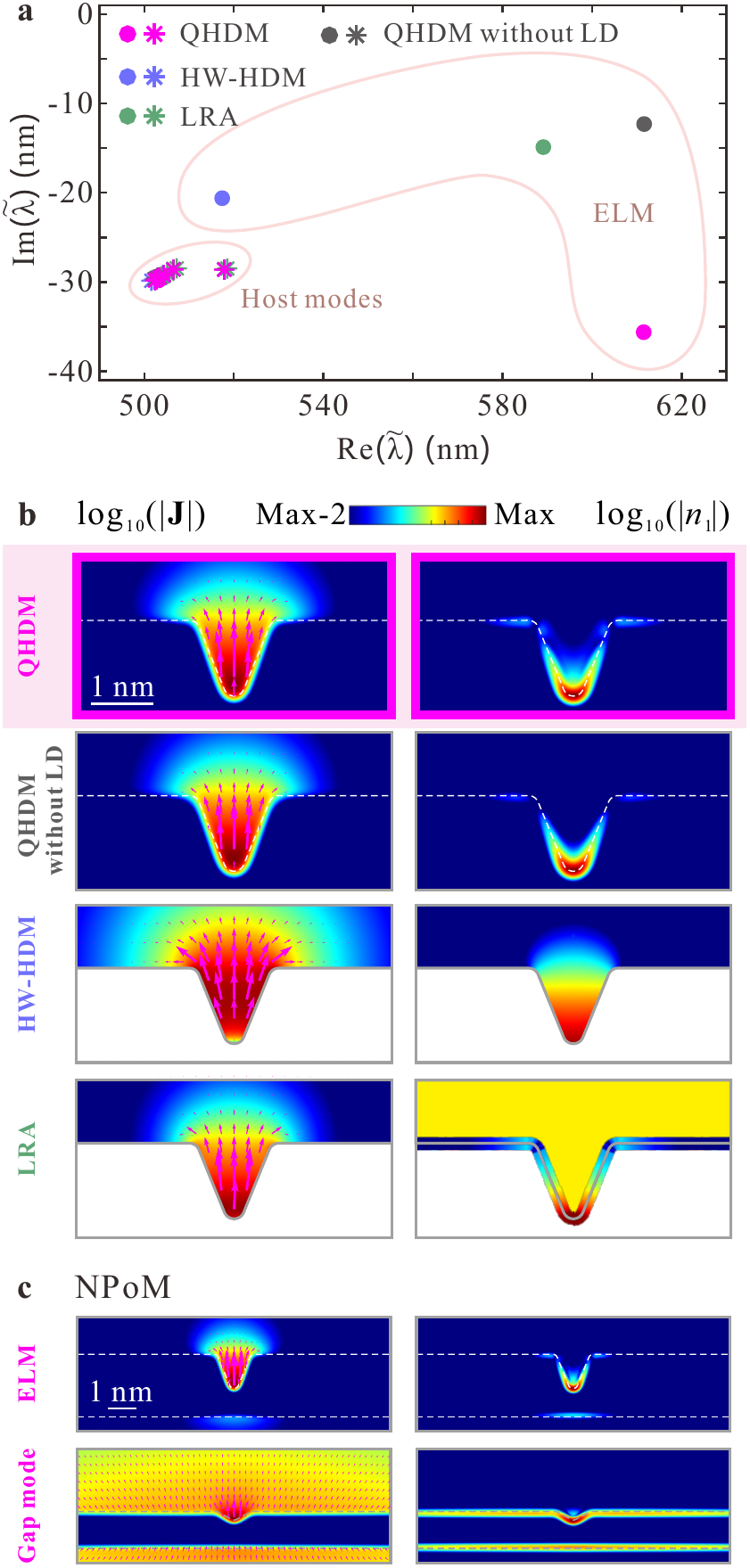}
\caption{\textbf{Quasi-normal mode (QNM) analysis for the protrusion-on-host structure.} \textbf{a,} Calculated eigen-wavelengths of the QNMs for the structure shown in Fig. \ref{Fig:1}\textbf{a} with $h_\mathrm{p} = 1.2$ nm using four different approaches. The eigen-wavelengths can be grouped into two categories, \textit{i.e.}, host modes and ELM. \textbf{b,} Distributions of the induced conduction current density $|\mathbf{J}|$ and electron density $|n_1|$ of the ELM in logarithmic scale calculated via different approaches. \textbf{c,} $|\mathbf{J}|$ and $|n_1|$ distributions of the ELM and the gap-plasmon mode studied in Fig. \ref{Fig:1}\textbf{h} and \ref{Fig:1}\textbf{g}, respectively. These calculations are according to the QHDM-based QNM formulation.}
\label{Fig:2}
\end{figure}

Next we expound the formation mechanism of ELMs by examining the on-resonance response of conduction electrons with methods at different levels of approximation in treating metal, namely, the classical local response approximation (LRA), the hard-wall hydrodynamic model\cite{Stephen2017NonlocalQNM} (HW-HDM, Supplementary Note 4) and QHDM with/without LD correction. 
Taking the $h_\mathrm{p}=1.2$ nm case in Fig. \ref{Fig:1}\textbf{a} as an example, we perform QNM analysis with the above methods and display the eigen-wavelengths in Fig. \ref{Fig:2}\textbf{a}. 
The existence of two kinds of modes, \textit{i.e.}, the host modes of the nanoparticle and the ELM, is evident. 
While all the methods predict the same eigen-wavelengths for the host modes, the incongruent predictions for the ELM allude to the criticality of the QHDM treatment. 
HW-HDM effectively causes electron spill-in and hence a blueshift of the ELM resonance by $94$ nm compared to the QHDM predictions, which correctly allow electrons to spill out of the metal boundary. 
LRA method, with no electron spill-out/in effect, produces a blueshift of $31$ nm with respect to the QHDM predictions. 
As expected, the effect of LD on the decay rate of ELM is significant and the quality factor of the ELM is reduced by $3$ times.

The induced current and electron density profiles of the ELM from all the methods are plotted in logarithmic scale in Fig. \ref{Fig:2}\textbf{b}. 
The common feature of these plots is the localization of the induced current and electron densities around the protrusion even though there is no inner physical boundary between the protrusion and host. 
Despite the common feature, the QHDM analysis gives the most rational distributions by allowing the induced current to fade away outside the geometry boundary. 
The electron density distributions exhibit greater difference. 
Under LRA, the induced charge density only exists on the surface 
In contrast, under HW-HDM, the electrons diffuse in the entire protrusion. 
The QHDM calculations reinstate the surface plasmon nature that the induced electrons reside in a thin layer and fade away from the boundary. 
LD then causes some charge diffusion in the protrusion, consistent with its charge diffusion interpretation\cite{Mortensen2014nonlocal}. 
One learns from these responses that driving the current and electrons around the protrusion is the pivotal factor in forming the ELM. 
Deeper distinction between the ELM and the gap-plasmon mode with the lightning rod effect is observed by comparing the response of conduction electrons. 
Figure \ref{Fig:2}\textbf{c} presents the induced current and electron density distributions for the ELM and the gap-plasmon mode studied in Fig. \ref{Fig:1}\textbf{h} and \ref{Fig:1}\textbf{g}, respectively. 
The distributions of the ELM are similar to the case without the gap as shown in Fig. \ref{Fig:2}\textbf{b}. 
In stark contrast, the current and electron densities of the gap mode delocalize over the metal around the gap despite observable influences from the lightning rod effect.

\begin{figure}[htpb]
\centering
\includegraphics[width=\linewidth]{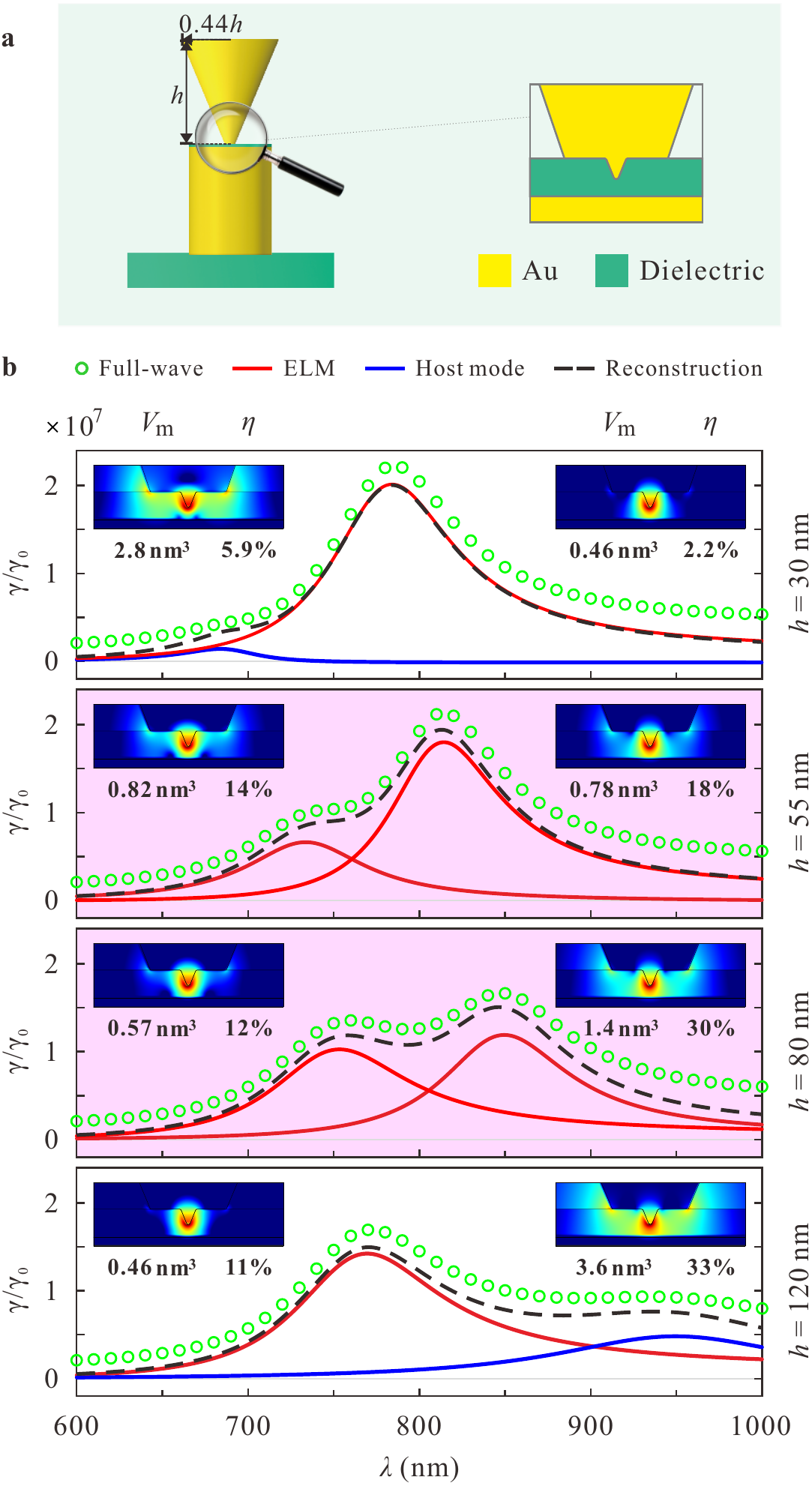}
\caption{\textbf{Radiation efficient ELM via hybridization with the host mode.} \textbf{a,} The schematic diagram of the nanocone-on-cylinder host antenna and the zoomed-in view of the $h_\mathrm{p} = 1.2$ nm protrusion in Fig. \ref{Fig:1}. \textbf{b,} Calculated emission enhancement spectra for the antennas with $h = 30$, $55$, $80$ and $120$ nm. The full-wave results, separate modal contributions from the ELMs and host modes, and the reconstruction are plotted with green circles, solid and black dashed traces, respectively. The insets display the mode profiles of the two QNMs with the mode volume and radiation efficiency indicated.}
\label{Fig:3}
\end{figure}

Extreme light concentration and high radiation efficiency are two ends difficult to reconcile\cite{Khurgin2017LandauDamping,Hulst2018PlasmonicCavityCoupling}, due to huge size mismatch with free-space wavelength and LD. 
Indeed, the radiative parts of the emission enhancement from the ELMs in Fig. \ref{Fig:1}\textbf{a} are barely over $0.001\%$. 
To solve the dilemma, we propose the idea of hybridizing the ELM with the host mode such that the radiation rate is greatly enhanced\cite{XWChen2012Metallodielectric} to outpace or compete with LD. 
We explore the idea by designing a gold nanocone host antenna for the $h_\mathrm{p}=1.2$ nm protrusion in Fig. \ref{Fig:1}\textbf{a}. 
As sketched in Fig. \ref{Fig:3}\textbf{a}, the size of the nanocone is characterized by its height $h$ and it forms a $2.2$ nm gap with a gold cylinder of $40$ nm radius and $105$ nm height. 
The emission enhancement spectra are calculated and plotted with green circles in Fig. \ref{Fig:3}\textbf{b} for antennas with $h=30$, $55$, $80$ and $120$ nm. 
The spectra are reconstructed with the modal contributions from the ELM and host mode (see Supplementary Note 5 for the remaining non-resonant contribution due to LD). 
The insets display the mode profiles of the two QNMs for each host antenna structure with the mode volume and radiation efficiency indicated. 
For $h=30$ and $120$ nm, the two QNMs are distantly separated in the spectra and the interaction is weak. 
For $h=55$ and $80$ nm, the interaction becomes strong and the consequences are that the ELM sacrifices a bit confinement whereas the host mode becomes extremely concentrated to be around 1 nm$^3$. 
Interestingly the radiation efficiencies of these hybridized ELMs are significantly boosted. 
Strikingly, for $h=80$ nm, the radiation efficiency of the main-contributing ELM with a mode volume of 1.4 nm$^3$ reaches $30\%$.
For $h=55$ nm, the emission enhancement around $808$ nm has a radiation efficiency of $18\%$ and is contributed vastly ($\sim 84\%$) by an ELM with a mode volume of only $0.78$ nm$^3$.

\begin{figure}[htpb]
\centering
\includegraphics[width=\linewidth]{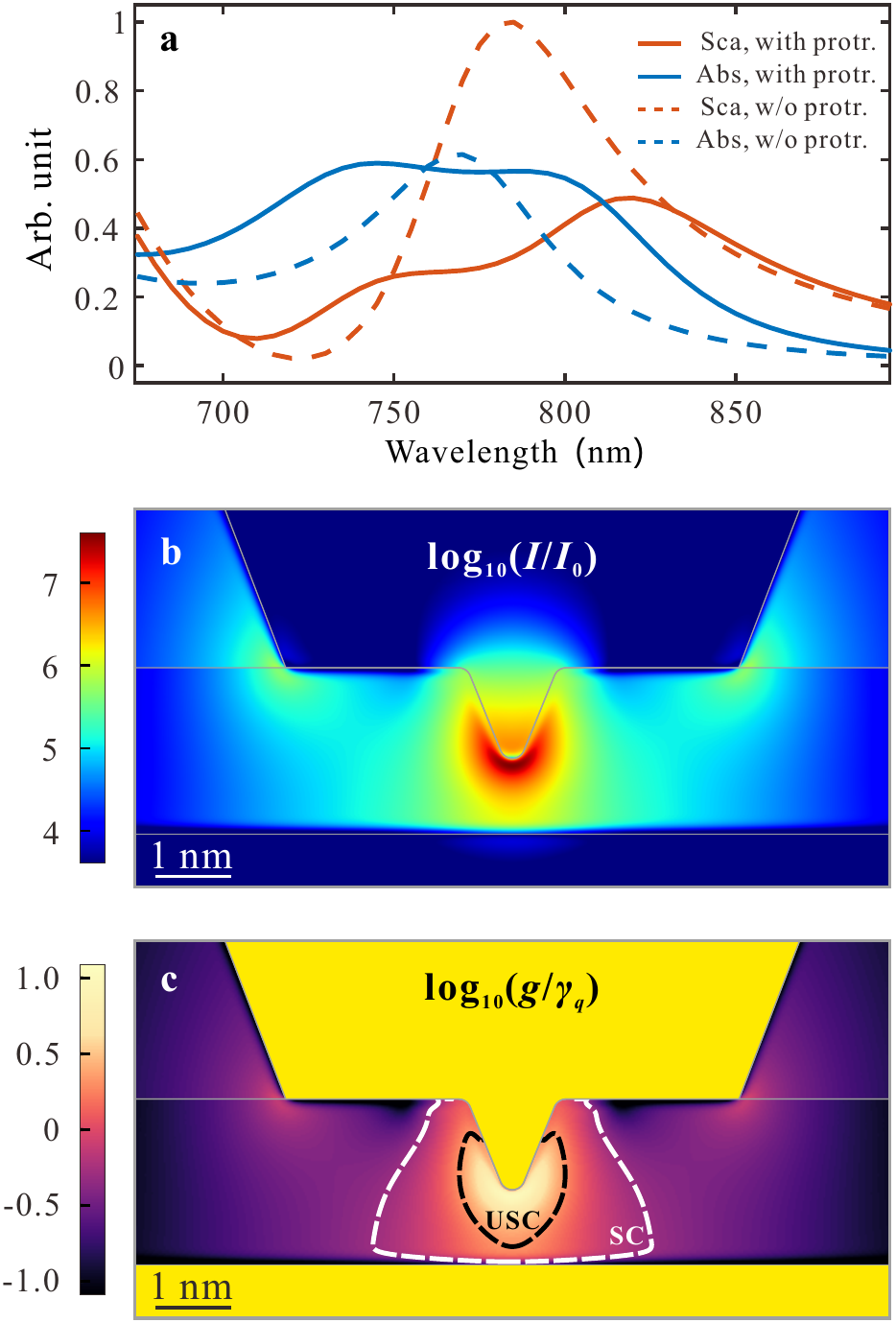}
\caption{\textbf{Far-field spectroscopy detection and excitation of the ELM and prospective ultrastrong light-matter interaction.} \textbf{a,} Calculated scattering and absorption spectra of antenna structure studied in Fig. \ref{Fig:3}\textbf{b} with $h = 55$ nm under the illumination of a focused radially polarized beam from the far field. Spectra without a protrusion are shown in dashed lines. \textbf{b,} Light intensity enhancement in logarithmic scale for the far-field excitation at $808$ nm. \textbf{c,} Atom-field coupling constant normalized by the ELM’s relaxation rate. The plot is in logarithmic scale and the white and black dashed contours demarcate the regions reaching strong and ultrastrong regimes with an optical dipolar emitter.}
\label{Fig:4}
\end{figure}

Efficient radiation from an ELM implies it could be perceived from far-field spectroscopy measurement. 
To exploit this possibility, we launch from the far field a focused radially polarized beam (RPB)\cite{Leuchs2003RadiallyPolarizedBeam} to illuminate the antenna structure shown in Fig. \ref{Fig:3}\textbf{b} with h = 55 nm (see Methods). 
The calculated scattering and absorption spectra of the antenna with and without the atomistic protrusion are shown in Fig. \ref{Fig:4}\textbf{a}. 
The effect of the protrusion is pronounced with both the scattering and absorption spectra split into two resonances corresponding to the two hybridized ELMs. 
This suggests an efficient delivery of the far-field optical energy into a tiny space at resonance. 
Figure \ref{Fig:4}\textbf{b} displays the distribution of the intensity enhancement with respect to the case without the antenna at the resonant wavelength of $808$ nm. 
The far-field excitation is effectively concentrated to a volume of ${\sim}1$ nm$^3$ with an unprecedented enhancement up to $4.0{\times}10^7$. 
Such tiny and intense hotspot has immediate implications for single-molecule surface-enhanced Raman scattering\cite{Zhang2013ChemicalMapping,Apkarian2019vibrationalmodes,Hongxing1999RamanScattering,LeRu2007RamanScattering,Jensen2017ACSNanoSingleMolecule,Stiles2008,ZhiYuan2020RamanSpectroscopy}. 
An optical mode with below 1 nm$^3$ volume promises an extremely strong interaction with a single emitter. 
The coupling strength with a dipolar emitter can be characterized by the coupling constant\cite{ExploringQuantum,Marquier2017Revisiting} $g=\sqrt{\gamma_0\gamma_q\rho(\mathbf{r},\omega_q)/2}$, with $\gamma_q=74$ meV being the decay rate of the ELM at $\omega_q=1.5$ eV and $\gamma_0=(10\,\mathrm{ns})^{-1}$ being the spontaneous decay rate of a typical emitter. 
The local density of states $\rho(\mathbf{r})$ can be approximated by the modal emission enhancement $f_{\mathrm{P},q}(\mathbf{r})$. 
Figure \ref{Fig:4}\textbf{c} maps the coupling constant. 
The white contour indicates the strong coupling threshold of $g=\gamma_q/2$ and the black contour denotes the ultrastrong coupling criterion of $g=0.1\omega_q$. 
The calculated maximum coupling energy under electric-dipole approximation reaches $790$ meV. 
The ultrastrong coupling phenomena with single emitters have mainly been discussed in the microwave range\cite{Nori2019UltrastrongCoupling} and now it seems possible with the ELMs at optical frequencies.

\section{Conclusions}

We have shown the robust existence of extremely localized modes around atomistic protrusions of a host nanoparticle. 
The predictions from our QHDM-based QNM analysis are reliable and tolerant to the changes of various approximations, including the von Weizs{\"a}cker parameter in $T_\mathrm{w}$ and correlation approximation (Supplementary Notes 6 and 7).
The counter-intuitive protrusion resonance is crucial for the extreme field localization.
By outpacing Landau damping through the hybridization with the host mode, we show that the originally dark extremely localized modes can be made bright with radiation efficiencies up to $30\%$, empowering the prospect of ultrafast and bright light sources, and huge enhancement of various low-cross-section processes such as Raman scattering\cite{LeRu2007RamanScattering} and nonlinear phenomena\cite{Zayats2016Nonlocality} under external far-field excitation. 
The findings associated with bright optical eigenmodes with about 1 nm$^3$ mode volume are extendable to other materials and geometries and underpin exciting new sciences, for instance, unlocking forbidden transitions\cite{Marin2016ShrinkingLight}, and atomic-scale single-molecule photochemistry\cite{Rubio2017StrongCouplingQED}.


\section{Methods}
\textbf{Treatment of bound electron response of gold.}
Absorption due to interband transitions in noble metals can be taken into account in the background permittivity $\varepsilon_\mathrm{b}(\widetilde{\omega})$ as in equation \eqref{Eq:2}\cite{Hongxing2015ResonanceShifts,Haus2014Tunneling}. 
In particular for gold, the background permittivity with the bound electron contribution can be extracted from the classical Drude-Lorentz model
\begin{align}
\varepsilon_\mathrm{r}
&= \left( \varepsilon_\infty 
+ \frac{\Delta\varepsilon_\mathrm{L}\omega_\mathrm{L}^2}{\omega_\mathrm{L}^2+2i\widetilde{\omega}\delta_\mathrm{L}-\widetilde{\omega}^2} 
- \frac{\omega_\mathrm{D}^2}{\widetilde{\omega}^2-i\widetilde{\omega}\gamma_\mathrm{D}} \right)
\notag\\
&\equiv
\frac{\varepsilon_\mathrm{b}(\widetilde{\omega})+\varepsilon_\mathrm{D}(\widetilde{\omega})}{\varepsilon_0}
\end{align}
The parameters $\varepsilon_\infty=5.90$, $\Delta\varepsilon_\mathrm{L}=1.27$, $\omega_\mathrm{L}=4.30{\times}10^{15}$ rad/s, $\delta_\mathrm{p}=3.74{\times}10^{14}$ rad/s, $\gamma_\mathrm{D}=4.11{\times}10^{13}$ rad/s and $\omega_\mathrm{D}=1.30{\times}10^{16}$ rad/s are determined by fitting the experimental data\cite{CRC}. 
As shown in Supplementary Fig. 1, the fitted and experimental complex permittivities agree very well in the wavelength range of interest ($480$ nm $\sim$ $1,000$ nm).

\vspace*{.5cm}
\textbf{Efficient FEM implementation of QHDM.} 
\textbf{a) 2D axisymmetric modeling.} 
The nanostructures under discussion bear cylindrical symmetry, which enables us to employ 2D axisymmetric modeling to ease the computational demand. 
Taking an isolated nanosphere with a protrusion as an example, the stationary and dynamic equations of QHDM only need to be solved on the cross section through the $z$ axis as shown in Supplementary Fig. 2\textbf{a}. 
In this cross sectional 2D model, the jellium metal boundary and the zoomed-in protrusion are depicted by black and golden lines, respectively. 
Conduction electrons typically spill $0.3$ nm $\sim$ $0.4$ nm over the jellium boundary. 
Thus we set the computation domain for $n_1$ as indicated by the blue line in the figure, $0.5$ nm beyond the jellium boundary, which is sufficiently large to obtain converged results. 
The computation domain for the stationary electron density $n_0$ is set $5$ nm beyond the jellium boundary, as denoted by the outmost red line in the figure, to correctly obtain the asymptotic distribution of $n_0$.

\textbf{b) Multiscale meshing.} 
A delicate meshing scheme is necessary to conduct accurate QHDM calculations for structures with the sizes of a few hundred nanometers and with sub-nanometer fine features. 
As shown in Supplementary Fig. 2b, we have applied a multiscale meshing scheme with the local mesh sizes denoted by $d_\mathrm{m}$. 
In particular, an extremely fine mesh with $d_\mathrm{m}\leq0.03$ nm is used in the immediate vicinity of the metal boundary to guarantee converged results and correctly capture the effects of nonlocality, spill-out and LD related to the characteristic Thomas-Fermi screening length $\lambda_\mathrm{TF}$ of ${\sim}0.1$ nm\cite{cirac2017CurrentDependent}. 
A mesh size of $d_m=0.5$ nm is assigned to the other metal area and near-field regions as indicated in Supplementary Fig. 2b. 
For the rest of the computation domain, a mesh size of $d_\mathrm{m}=20$ nm suffices. 
The meshes near the protrusion and metal boundary are showcased in Supplementary Fig. 2c and 2d. 
All the QHDM and QNM calculations are performed in a work station with $25$ processors ($2.50$ GHz) and $256$ GB memory.

\vspace*{.5cm}
\textbf{ELM excitation by a focused RPB.} 
As a far-field excitation, a focused RPB illuminates upwards from the dielectric substrate on the antenna structure. 
The electric field of the RPB takes the form\cite{Leuchs2001RPB}
\begin{align*}
\begin{pmatrix}
E_r \\ E_z
\end{pmatrix}
=
k\int_0^{\theta}d\alpha\,A(\alpha)\sin\alpha\,e^{-ik_0z\cos\alpha}
\\ \times
\begin{pmatrix}
\cos\alpha\,J_1(-kr\sin\alpha)\\
i\sin\alpha\,J_0(-kr\sin\alpha)
\end{pmatrix}
\end{align*}
where $A(\alpha)=f_l\sin\alpha\,e^{-(f\sin\alpha)^2/\omega_0^2}\sqrt{|\cos\alpha|^2}$. 
$k=1.45k_0$ is the wavevector in the silica substrate. 
The focal length $f_l=1.8$ mm, the beam radius $\omega_0=0.82$ mm and the maximum angle of incidence of the plane waves $\theta=90^{\circ}$ are used in the calculations\cite{XWChen2010RPB}. 
The incident focused RPB is first simulated in silica to obtain the electromagnetic fields $(\mathbf{E}_\mathrm{inc}, \mathbf{H}_\mathrm{inc})$. 
The electric field distribution $\mathbf{E}_\mathrm{inc}$ is shown in Supplementary Fig. 4a. 
The equivalent current sources
\begin{equation}
\mathbf{J_s}
= -H_{\mathrm{inc},\phi}\,\mathrm{\widehat{\mathbf{r}}},\qquad
\mathbf{M_s}
= -E_{\mathrm{inc},r}\,\mathrm{\widehat{\mathbf{z}}}
\end{equation}
are then set on a horizontal surface far below the antenna structure to simulate far-field excitation in the following calculations. 
To find the reference for computing intensity enhancement factor, the focused RPB is next incident on the bare silica-vacuum interface. 
The resulting electric field $|\mathbf{E}_\mathrm{ref}|$ is illustrated in Supplementary Fig. 4b. 
At last, the antenna on the silica interface is excited with the focused RPB, leading to the electric field $\mathbf{E}_\mathrm{exc}$. 
The distribution of the intensity enhancement factor is computed as $|\mathbf{E}_\mathrm{exc}/\mathbf{E}_\mathrm{ref}|^2$.

\section{Data availability}
The data that support the plots within this paper and other findings of this study are available from the corresponding author upon reasonable request.

\section*{Acknowledgments}
We gratefully acknowledge financial support from the National Natural Science Foundation of China (Grant Number 11874166, 11604109) and Huazhong University of Science and Technology.


\section*{Author contributions}
X.-W.C conceived the concepts behind this research. X.-W.C and P.Z. designed and supervised the research project. P.Z., X.-W.C, W.L. and Q.Z. developed theoretical formulation, analysis and interpretation. W.L. and Q.Z. performed the numerical calculations. X.-W.C. and P. Z. wrote the manuscript with inputs from W.L. and Q.Z.

\section*{Competing interests}
The authors declare no competing interests.

\section*{Additional information}

\textbf{Correspondence and requests for materials} should be addressed to X.-W.C or P.Z.



\bibliography{QHDM_QNM_arXiv_ref}


\end{document}